\documentclass[aps,pra,reprint,superscriptaddress]{revtex4-1}

\usepackage{amsmath}    
\usepackage{graphicx}   
\usepackage[lofdepth,lotdepth, caption=false]{subfig}
\usepackage{verbatim}   
\usepackage{color}      
\usepackage{hyperref}   
\usepackage{dcolumn}
\usepackage{float}

\begin{document}

\title{Substitution of Ni for Fe in superconducting Fe$_{0.98}$Te$_{0.5}$Se$_{0.5}$ depresses the normal-state conductivity but not the magnetic spectral weight}
\author{Jinghui Wang}
\altaffiliation{These authors contributed equally to the work.}
\affiliation{Center for Superconducting Physics and Materials,
National Laboratory of Solid State Microstructures, Department of Physics,
and Collaborative Innovation Center of Advanced Microstructures, Nanjing University, Nanjing 210093, China}
\author{Ruidan Zhong}
\altaffiliation{These authors contributed equally to the work.}
\affiliation{Condensed Matter Physics and Materials Science Department, Brookhaven National Laboratory, Upton, NY 11973, USA}
\affiliation{Materials Science and Engineering Department, Stony Brook University, Stony Brook, NY 11794, USA}
\author{Shichao Li}
\author{Yuan Gan}
\affiliation{Center for Superconducting Physics and Materials,
National Laboratory of Solid State Microstructures, Department of Physics,
and Collaborative Innovation Center of Advanced Microstructures, Nanjing University, Nanjing 210093, China}
\author{Zhijun~Xu}
\affiliation{Physics Department, University of California, Berkeley, CA 94720, USA}
\affiliation{Materials Science Division, Lawrence Berkeley National Laboratory, Berkeley, CA 94720, USA}
\author{Cheng~Zhang}
\affiliation{Condensed Matter Physics and Materials Science Department, Brookhaven National Laboratory, Upton, NY 11973, USA}
\affiliation{Materials Science and Engineering Department, Stony Brook University, Stony Brook, NY 11794, USA}
\author{T.~Ozaki}
\affiliation{Condensed Matter Physics and Materials Science Department, Brookhaven National Laboratory, Upton, NY 11973, USA}
\author{M.~Matsuda}
\affiliation{Quantum Condensed Matter Division, Oak Ridge National Laboratory, Oak Ridge, TN 37831, USA}
\author{Yang Zhao}
\affiliation{NIST Center for Neutron Research, National Institute of Standards and Technology, Gaithersburg, MD 20899, USA}
\affiliation{Department of Materials Science and Engineering, University of Maryland, College Park, MD 20742, USA}
\author{Qiang Li}
\author{Guangyong~Xu}
\author{Genda Gu}
\author{J.~M.~Tranquada}
\affiliation{Condensed Matter Physics and Materials Science Department, Brookhaven National Laboratory, Upton, NY 11973, USA}
\author{R.~J.~Birgeneau}
\affiliation{Physics Department, University of California, Berkeley, CA 94720, USA}
\affiliation{Materials Science Division, Lawrence Berkeley National Laboratory, Berkeley, CA 94720, USA}
\author{Jinsheng Wen}
\altaffiliation{jwen@nju.edu.cn}
\affiliation{Center for Superconducting Physics and Materials,
National Laboratory of Solid State Microstructures, Department of Physics,
and Collaborative Innovation Center of Advanced Microstructures, Nanjing University, Nanjing 210093, China}

\date{\today}

\begin{abstract}
We have performed systematic resistivity and inelastic neutron scattering measurements on Fe$_{0.98-z}$Ni$_z$Te$_{0.5}$Se$_{0.5}$ samples to study the impact of Ni substitution on the transport properties and the low-energy ($\le$~12~meV) magnetic excitations. It is found that, with increasing Ni doping, both the conductivity and superconductivity are gradually suppressed; in contrast, the low-energy magnetic spectral weight changes little.  Comparing with the impact of Co and Cu substitution, we find that the effects on conductivity and superconductivity for the same degree of substitution grow systematically as the atomic number of the substituent deviates from that of Fe.  The impact of the substituents as scattering centers appears to be greater than any contribution to carrier concentration.  The fact that low-energy magnetic spectral weight is not reduced by increased electron scattering indicates that the existence of antiferromagnetic correlations does not depend on electronic states close to the Fermi energy.
\end{abstract}


\maketitle

\section{Introduction}
Research on the Fe-based superconductors has been intense since the initial discovery of superconductivity in LaO$_{1-x}$F$_{x}$FeAs (labelled 1111 according to the stoichiometry of the parent compound) with $T_{c}=26$~K \cite{Kamihara2008}.  Among the studied systems, the phase diagrams found for Ba(Fe$_{1-x}$TM$_x)_2$As$_2$ (labelled 122, with TM = 3$d$ transition metals such as Co and Ni) have been quite intriguing \cite{Sefat2008,1367-2630-11-2-025008}, as the TM is isovalent to Fe, whereas in the 1111 \cite{Kamihara2008} and Ba$_{1-x}$(Na/K)$_{x}$Fe$_{2}$As$_{2}$ \cite{Rotter2008,doi:10.1021/cm100956k} cases, superconductivity is achieved by heterovalent doping. Initially, the tuning of superconductivity in the 122 system by substitution of Co or Ni had been understood by assuming that the TM ions simply contributed their extra electrons to the conduction bands \cite{canfield:060501,PhysRevB.82.024519,PhysRevB.84.020509}, with a resulting rigid-band shift of the Fermi level \cite{PhysRevB.83.094522,PhysRevB.83.144512,JPSJ.80.123701,Kemper2010}.  However, theoretical analyses have indicated that the extra electrons of the TM dopants do not entirely delocalize and that at least part of the doping effect is associated with impurity scattering \cite{PhysRevLett.105.157004,2011arXiv1112.4858B,Vavilov2011}. These proposals have been supported by spectroscopic studies \cite{PhysRevLett.107.267402,PhysRevLett.109.077001,PhysRevB.86.104503,PhysRevLett.110.107007,0953-8984-24-21-215501}, as well as by studies of magnetic correlations \cite{PhysRevLett.109.167003,PhysRevLett.113.117001}.   One also finds that the dependence of the superconducting dome as a function of dopant concentration does not follow the scaling behavior predicted by the rigid-band model  \cite{Sefat2008,1367-2630-11-2-025008,canfield:060501,PhysRevB.84.054540,PhysRevB.82.024519}.                   

In the system Fe$_{1+y}$Se, superconductivity occurs in the nearly-stoichiometric compound without a need to overcome antiferromagnetic order \cite{mkwreview1,mcqueen:014522,Hu2011}.   It is possible to enhance $T_c$ by partial substitution of Te for Se \cite{kata10,liupi0topp}; however, substitution of Co, Ni, or Cu for Fe in Fe$_{1+y}$Te$_{1-x}$Se$_{x}$, inevitably leads to a reduction in $T_c$ \cite{danielfeseco,williams-2009-21,PhysRevB.82.104502,mkwreview1,2010arXiv1010.4217G,fetesenico1}.  For a given TM-dopant concentration, the depression of $T_c$ grows as one moves from Co to Ni to Cu.  In previous work, we have confirmed that Cu reduces $T_c$ rapidly; furthermore, with 10\% Cu substitution in Fe$_{0.98-z}$Cu$_z$Te$_{0.5}$Se$_{0.5}$, the resistivity has the temperature dependence of an insulator \cite{PhysRevB.88.144509}.  We also observed that low-energy ($\leq 12$~meV) antiferromagnetic spectral weight is significantly enhanced with Cu doping, but without inducing order.  The impact of Cu on resistivity suggests strong scattering by the dopants, resulting in localization of conduction electrons, and it is not surprising that this would lead to the destruction of superconductivity.  On the other hand, the Cu does not depress the magnetic correlations, which are believed to be important to superconductivity.  In fact, the effect on the magnetism says something significant about the interactions responsible for the antiferromagnetic correlations.  Now, Cu has the most extreme impact on the electronic transport, but is it qualitatively different from that induced by Ni or Co dopants?

In this paper, we attempt to answer this question by performing systematic resistivity and inelastic neutron scattering measurements on Fe$_{0.98-z}$Ni$_{z}$Te$_{0.5}$Se$_{0.5}$ single-crystal samples, with $z=0.02$, 0.04, 0.10 (labelled as Ni02, Ni04, and Ni10 respectively). The results are discussed with reference to those of the Cu-doped case \cite{PhysRevB.88.144509}. With increasing Ni content, $T_c$ is gradually suppressed. With 10\% Ni doping, the resistivity increases slowly with decreasing temperature, exhibiting a weakly insulating behavior. The low-energy magnetic correlations are modified somewhat by the Ni substitution, but the magnetic spectral weight in the normal state changes relatively little. Compared with the results from the Cu-substituted samples, these impacts of Ni doping are reduced in magnitude but qualitatively similar. Our results are compatible with the theoretical arguments that disorder and scattering are significant consequences of TM substitution \cite{PhysRevLett.105.157004,2011arXiv1112.4858B}; in addition, the interactions responsible for magnetic correlations must be short range.

\section{Experimental}
Single-crystal samples of Fe$_{0.98-z}$Ni$_{z}$Te$_{0.5}$Se$_{0.5}$ with nominal concentrations of $z$=0.02, 0.04, and 0.10 (labelled as Ni02, Ni04, and Ni10 respectively) were grown by the horizontal Bridgman method \cite{interplaywen}.  To start, the raw materials (99.999\% Te, 99.999\% Se, 99.99\% Fe, and 99.99\% Ni) were weighed and mixed with the desired molar ratio, and then doubly sealed into evacuated high-purity (99.995\%) quartz tubes. The materials were put into the furnace horizontally and heated in the following sequence: ramped to 660~$^{\circ}$C in 3~h; held for 1~h; ramped to 900~$^{\circ}$C in 2~h; held for 1~h; ramped to 1000~$^{\circ}$C in 1~h; held for 12~h; cooled to 300~$^{\circ}$C with a cooling rate of $-0.5$ or $-1^{\circ}$C~h$^{-1}$; then the furnace was shut down and cooled to room temperature. There was a small temperature gradient in the furnace from one end to the other, so that the melted liquid crystallized unidirectionally. To minimize the effects of Fe interstitials, we used a nominal Fe composition of 0.98 instead of 1 for all samples. From X-ray and neutron powder diffraction, and inductively coupled plasma (ICP) measurements on the sample compositions, the maximum deviation of the real composition from the nominal one was determined to be less than 2\% \cite{xudoping11,2011arXiv1108.5968Z}. The $a$-$b$ plane resistivity was measured using a four-point configuration with four contacts made on the $a$-$b$ plane, in a commercial cryostat, with an applied current of 5~mA. The samples used in the resistivity measurements were cut from the same respective batches used in the neutron scattering measurements. The typical dimension was $7\times2\times0.4$~mm$^3$.

Neutron scattering experiments on Ni02 and Ni10 were carried out on the BT-7 triple-axis spectrometer at the NIST Center for Neutron Research, using a beam collimating configuration of Open-$80'$-Sample-$80'$-$120'$. Two pyrolytic graphite (PG) filters were placed after the sample to reduce contamination from higher-order neutrons. The final energy, $E_f$, was fixed at 14.7~meV. The Ni04 and Ni10 (the same sample measured on BT-7) samples were measured on the HB1 triple-axis spectrometer at the High Flux Isotope Reactor, Oak Ridge National Laboratory. The beam collimations were $48'$-$40'$-Sample-$40'$-$240'$ with 2 PG filters after the sample. The $E_f$ was fixed at 13.5~meV.  Each of the crystals was a semicylinder, with two flat cleavage surfaces and a mass larger than 10~g. The crystals were mounted in aluminum sample holders and loaded into a closed-cycle refrigerator (CCR). The experiments were performed in the $(HK0)$ plane defined by the [100] and [010] wave vectors. The wave vectors, ${\bf Q}$, will be expressed in terms of reciprocal lattice units (rlu) of $(a^*, b^*,c^*)=(2\pi/a,2\pi/b,2\pi/c)$, where the room-temperature lattice constants are $a=b\approx3.8$~\AA, and $c=6.1$~\AA, corresponding to a unit cell with two Fe atoms.   The measured intensity $I_{\rm meas}$ was converted to the dynamical spin correlation function $S({\bf Q},E)$ with absolute unit of $\mu_{\rm B}^2$eV$^{-1}$/Fe by the integrated incoherent elastic scattering intensity ${I_{\rm inc}}$ measured at  (0.7, 0.3, 0), and (0.7, 0.7, 0) and averaged, using the formula \cite{neutron1,gynormal}
\begin{equation*}
S({\bf Q},E)=\frac{I_{\rm meas}\mu^2_{\rm B}}{4\pi |f({\bf Q})|^2 p^2}\cdot  \frac{\sum_jn_j\sigma_{{\rm inc}, j}}{I_{\rm inc}},
\end{equation*}
where $\mu_{\rm B}$ is the Bohr magneton, $f(\bf Q)$ is the magnetic form factor of Fe$^{2+}$, $p = 0.27\times10^{-12}$~cm, $n_j$ and $\sigma_{{\rm inc}, j}$ are the molar ratio and the incoherent cross section for the element $j$ in the compound, respectively. 

\section{Results}

\begin{figure}[htb]
\begin{center}
\includegraphics[width=0.9\linewidth] 
{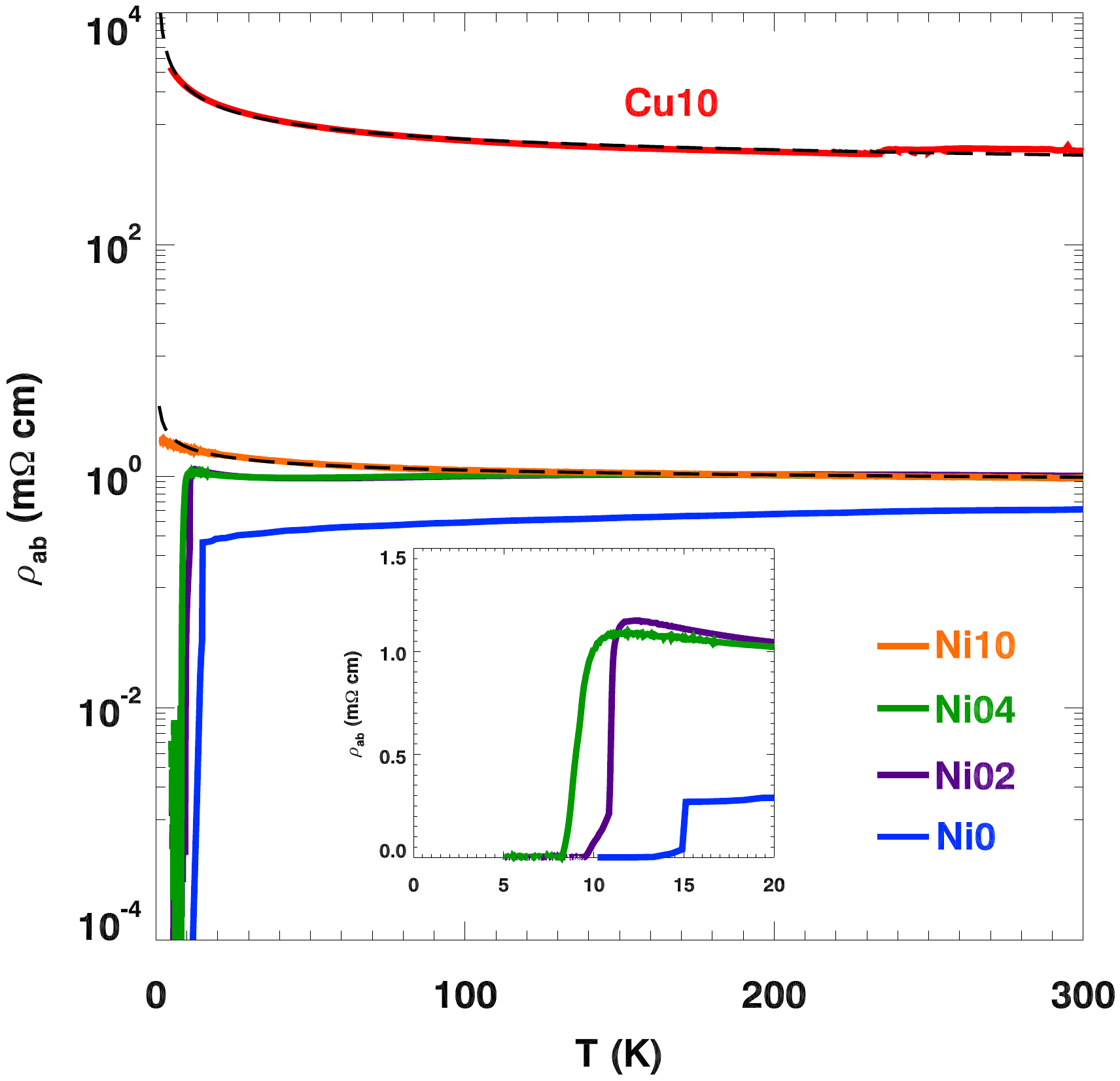}
\caption{\label{fig:rt1}(Color online) $a$-$b$ plane resistivity ($\rho_{ab}$) for Ni0, Ni02, Ni04, Ni10, and Cu10 in the semi-log scale. Dashed lines are results of fits to the data with the three-dimensional Mott variable range hopping formula, as described in the text. Inset shows $\rho_{ab}$ vs temperature in the low-temperature range for the three superconducting samples, Ni0, Ni02 and Ni04.}
\end{center}
\end{figure}

\begin{figure}[htb]
\begin{center}
\includegraphics[height=0.9\linewidth,angle=90] 
{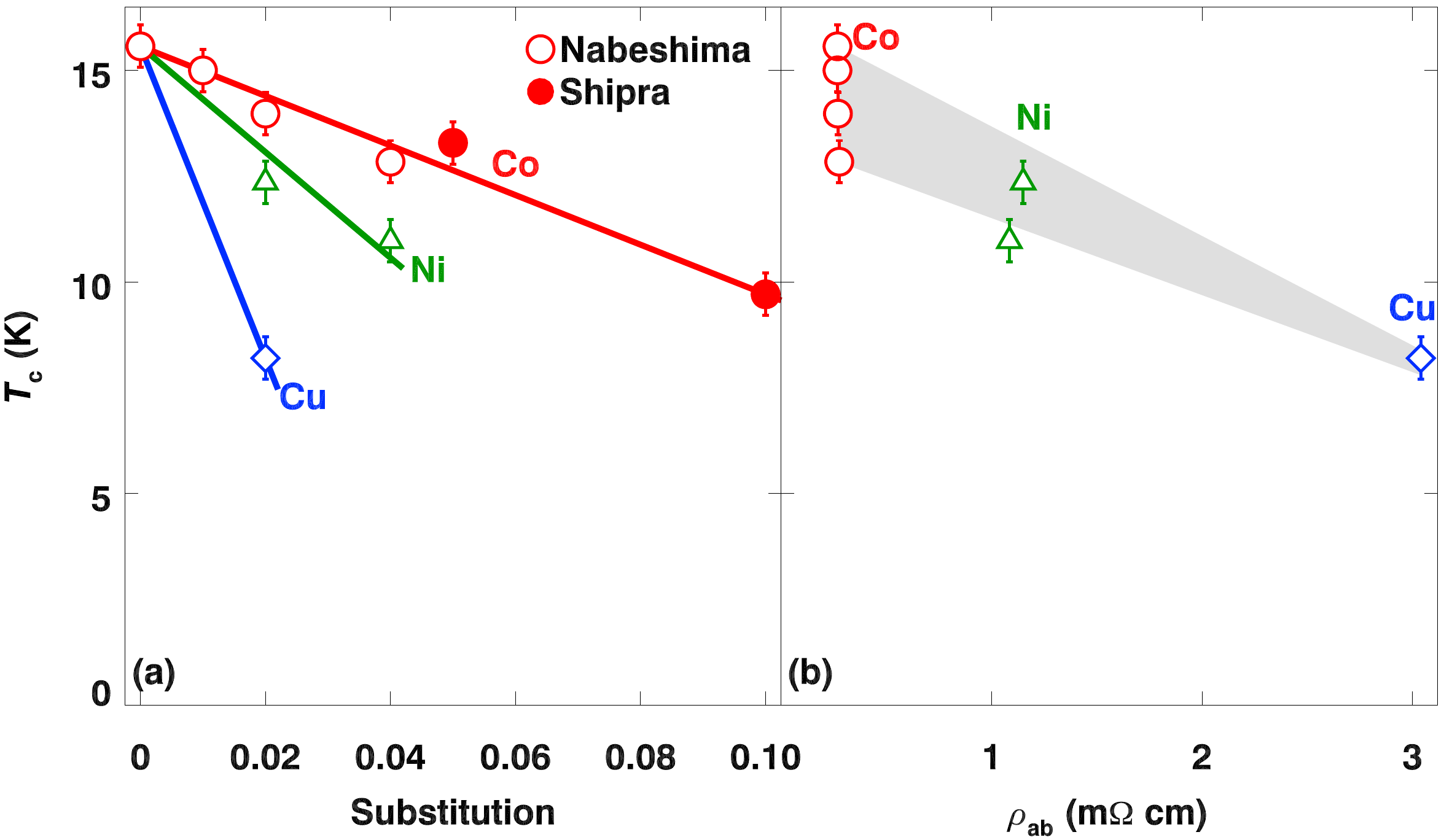}
\caption{\label{fig:tczrho}(Color online) $T_c$ as a function of TM concentration (a), and resistivity at the temperature where $\rho_{ab}$ starts to drop (b) for Co, Ni and Cu. Open and closed circles are data extracted from the work by Nabeshima {\it et al.},~\cite{1347-4065-51-1R-010102} and Shipra {\it et al.}~\cite{fetesenico1} respectively. Lines in (a) and shade in (b) are guides to the eyes. Error represents one standard deviation $\sigma$ throughout the whole paper.}
\end{center}
\end{figure}

\begin{figure*}[htb]
\begin{center}
\includegraphics[width=0.8\linewidth]
{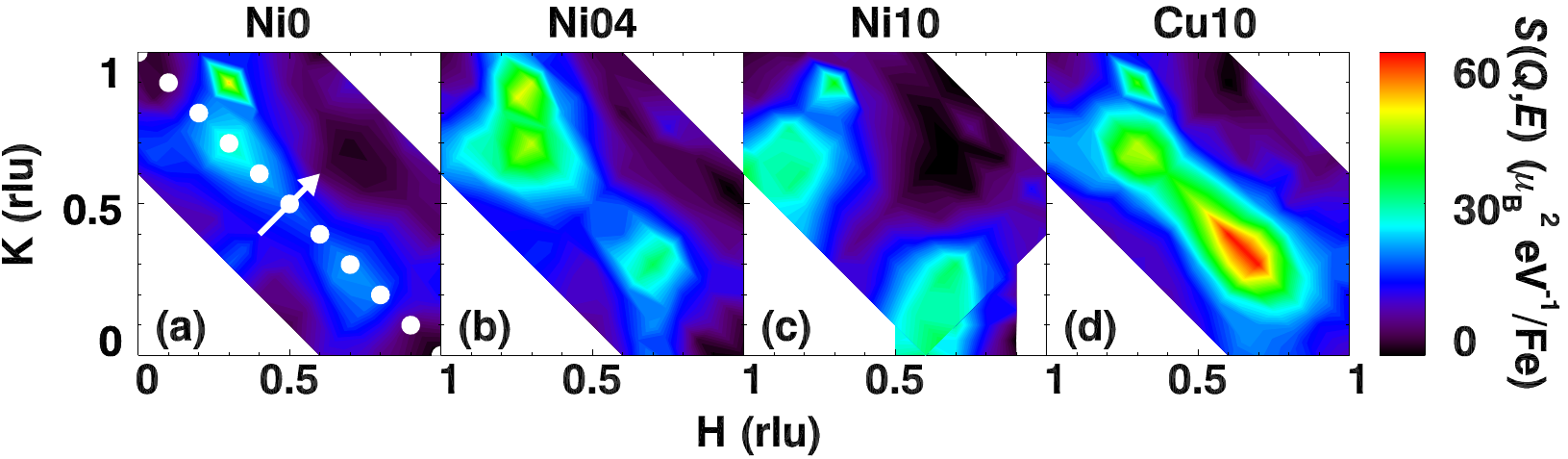}
\caption{\label{fig:meshni7} (Color online) Contour plots of the magnetic scattering at a constant energy of 6~meV at 100~K for Ni0, Ni04, Ni10, and Cu10. The data are obtained by performing a series of linear scans along the [110] direction (illustrated by the arrow) through the positions indicated by the dots in (a). The bright spot close to (0.3, 1) is a spurion.}
\end{center}
\end{figure*}

\begin{figure}[htb]
\begin{center}
\includegraphics[width=0.9\linewidth]
{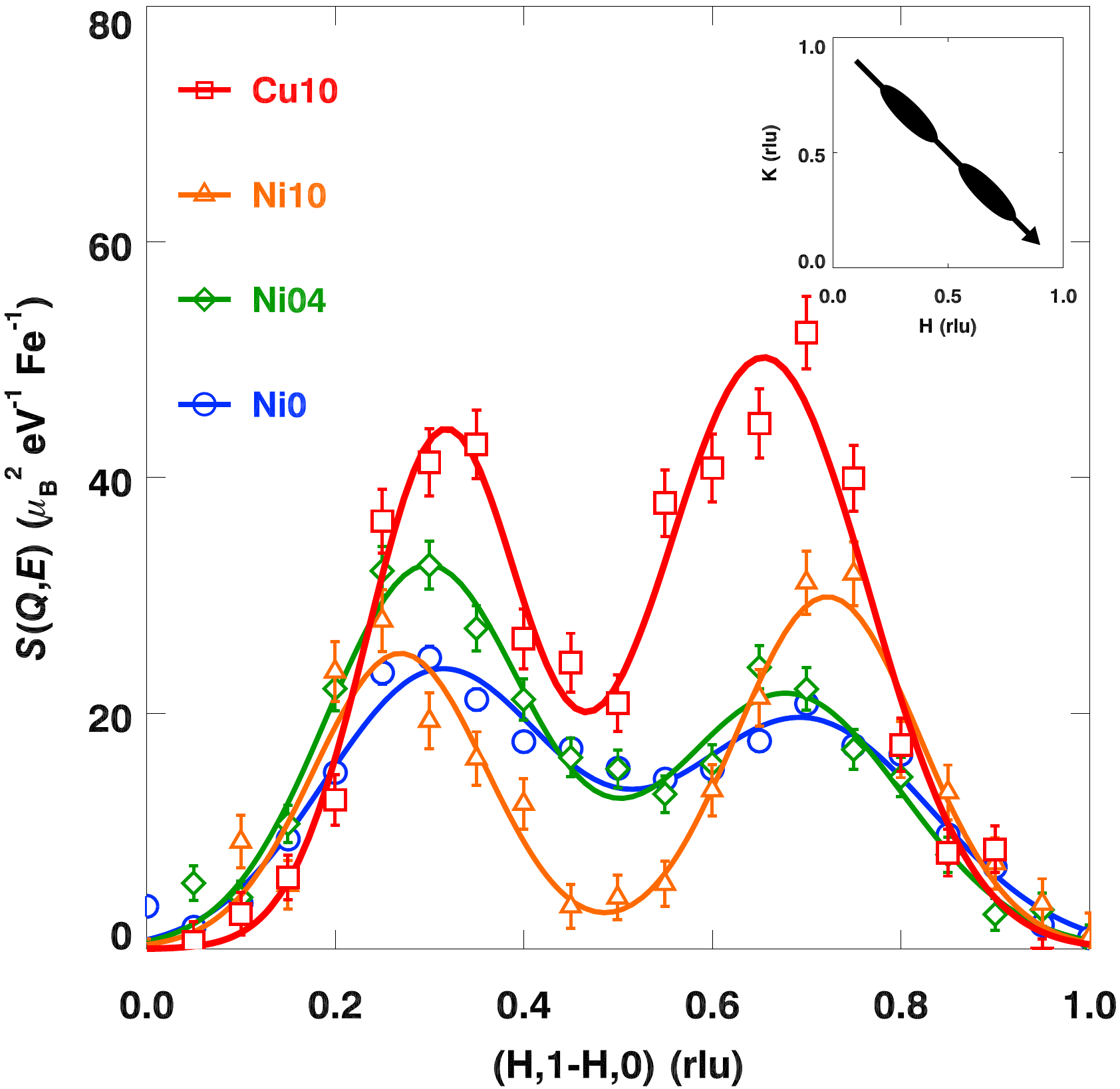}
\caption{\label{fig:trans} (Color online) Linear scans through (0.5,\,0.5) along the [1\={1}0] direction with an energy transfer of 6~meV at 100~K for Ni0, Ni04, Ni10, and Cu10. Solid lines through the data are the results of fits with double Gaussians.}
\end{center}
\end{figure}

\begin{figure}[htb]
\begin{center}
\includegraphics[width=0.9\linewidth]
{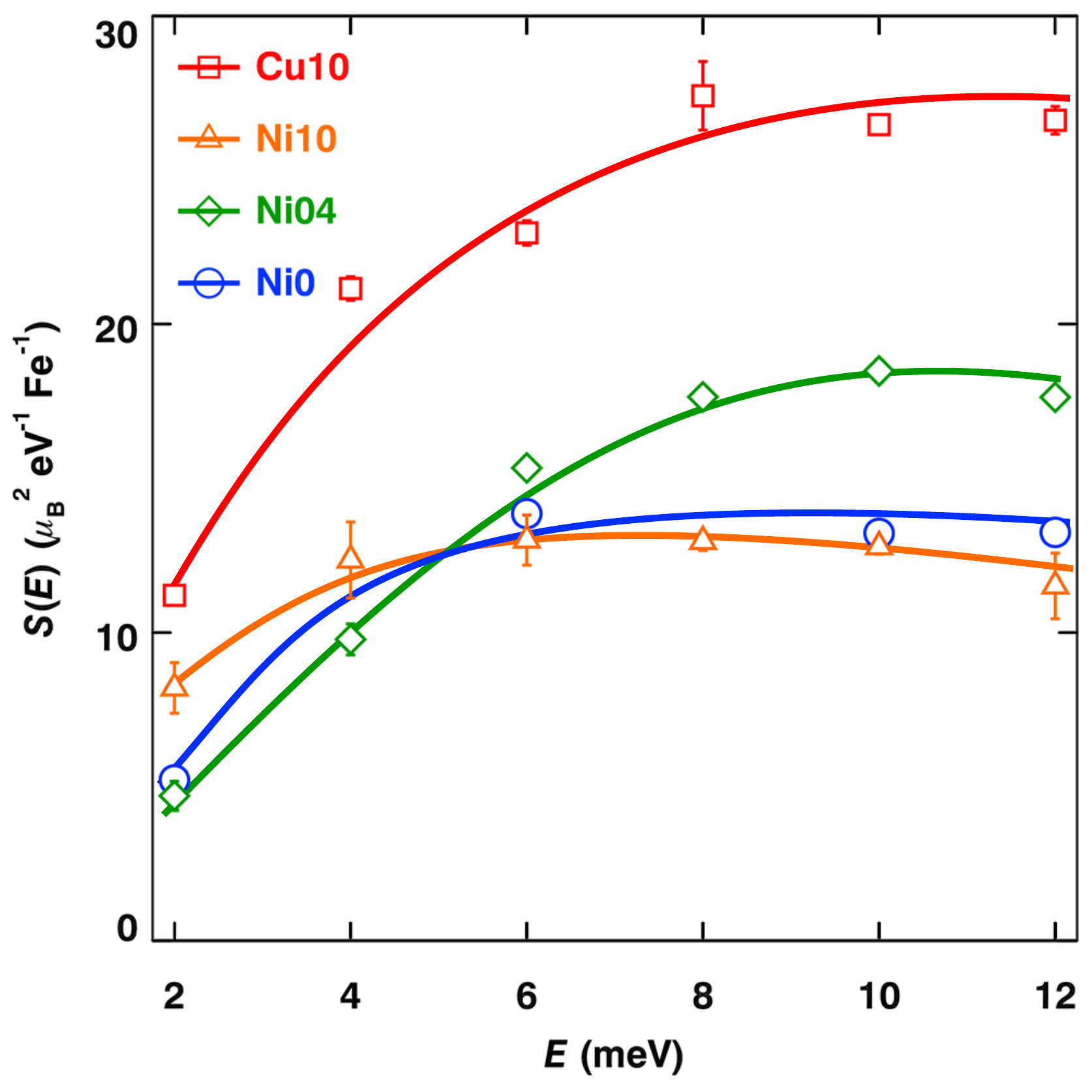}
\caption{\label{fig:se} (Color online) Q-integrated intensities of the constant-energy scans shown in Fig.~\ref{fig:trans}, but at energies ranging from 2 to 12~meV with a 2-meV interval at 100~K for Ni0, Ni04, Ni10, and Cu10. Solid lines are guides to the eyes.}
\end{center}
\end{figure}

Resistivity data with the current running in the \emph{a-b} plane, $\rho_{ab}$ for Ni02, Ni04, and Ni10 are shown in Fig.~\ref{fig:rt1}. For comparison, the resistivity for a Ni-free sample, Ni0, and 10\% Cu-substituted sample, Fe$_{0.88}$Cu$_{0.1}$Te$_{0.5}$Se$_{0.5}$ (Cu10), are also plotted. With Ni doping, superconductivity is gradually suppressed. For Ni02, the resistivity starts to drop at $\sim$~12~K, and zero resistivity is reached at $\sim$~9.8~K, as shown in the inset of Fig.~\ref{fig:rt1}; these two temperatures for the Ni04 sample are 10.5~K and 8.5~K respectively. The Ni10 is not superconducting down to the lowest temperature measured (2~K). The absolute values of the resistivity in the normal state are also higher in the Ni-substituted samples, indicating a suppression of the electrical conductivity.

Compared with Cu, the impact of Ni substitution on the normal-state resistivity is much more benign \cite{PhysRevB.88.144509}. Ni increases the normal-state resistivity with respect to the Ni-free sample by roughly a factor of 4 to 5 at 20~K, whereas the increase is almost 4 orders of magnitude for the Cu10 sample. The latter resistivity  can be fitted rather well with a three-dimensional Mott variable-range-hopping formula \cite{Mott1969} $\rho_{ab}=\rho_0{\rm exp(T_0}/T^{1/(1+d)})$, as indicated in Fig.~\ref{fig:rt1} by the dashed line following the Cu10 data; here, $\rho_0$ and $T_0$ are constants, and $d=3$ is the dimensionality. This indicates that the Cu10 sample behaves like a three-dimensional Mott insulator.   A similar fit to the Ni10 data works over a substantial temperature range, but overshoots at low temperature, perhaps due to the presence of residual superconducting correlations.

In Fig.~\ref{fig:tczrho}(a) we plot $T_c$ (onset) as a function of the TM substitution $z$ for our Ni and Cu-doped samples and compare with results for Co-doping (in samples with similar Te concentration) from Refs.~\onlinecite{1347-4065-51-1R-010102,fetesenico1}.  One can see that there is a monotonic increase in the rate of $T_c$ suppression in going from Co to Ni to Cu. For Co, the sample is superconducting with $z$ up to 10\%, where $T_c$ is still close to 10~K \cite{fetesenico1}. For Ni, $T_c$ drops to zero somewhere between 4\% and 10\% doping, while for Cu, the cutoff $z$ for superconductivity is $\sim2$\%. For Co, Ni and Cu, the $T_c$ reduction rates are $-0.58$, $-1.24$, and $-3.68$~K per 1\% substitution, respectively. In Fig.~\ref{fig:tczrho}(b), it is shown that $T_c$ is anticorrelated with the normal-sate resistivity. For Co doping, where the resistivity is quite close to that of the undoped sample, $T_c$ is higher; Ni lies in the intermediate range between Co and Cu. The rate of $T_c$ reduction tends to follow the normal-state resistivity. 

Next, we turn to the inelastic neutron scattering measurements of the magnetic excitations. We have performed a series of scans around the two characteristic wave vectors $(0.5,0)$ and $(0.5,0.5)$ at excitation energies of 2 to 12~meV, with a 2-meV interval. As in previous studies for Se contents close to 50\% \cite{PhysRevLett.109.227002,xudoping11,lumsden-2009,liupi0topp,PhysRevB.87.224410,1367-2630-14-7-073025,PhysRevB.88.144509}, there is little spectral weight around $(0.5,0)$; hence, we will focus on the results near $(0.5,0.5)$.  While we have carried out measurements over a temperature range from 5 to 300~K, the trends in magnetic spectral weight with doping are adequately captured by considering just the data obtained at 100~K (well above the maximal $T_c$ of this system to avoid any effects from the superconducting correlations). Furthermore, the results for Ni02 appear to be very similar to those of the Ni04 sample. Taking all these  points into account, we plot in Fig.~\ref{fig:meshni7} representative results as contour maps around $(0.5,0.5)$ at a constant energy of 6~meV for each of the Ni0, Ni04, and Ni10 samples. These maps are obtained by plotting a series of linear scans with the trajectories shown in Fig.~\ref{fig:meshni7}(a). For comparison, the data for Cu10 are also shown. At this temperature of 100~K, the magnetic scattering peaks at wave vectors displaced from $(0.5,0.5)$ along the $[1\bar{1}0]$ direction, as in previous work \cite{PhysRevLett.109.227002,PhysRevB.88.144509,PhysRevB.89.174517}. 

To provide a better comparison of the strength of the magnetic excitations, in Fig.~\ref{fig:trans}, we plot linear scans through $(0.5,0.5)$ along the $[1\bar{1}0]$ direction with an energy transfer of 6~meV at 100~K for the Ni0, Ni04, Ni10 and Cu10 samples.  While there is some variation in the wave vectors of the peaks, the widths stay relatively constant. To compare the overall spectral weight at low energies, we have {\bf Q}-integrated the intensities for scans from 2 to 12 meV; the results are plotted as a function of the energy transfer in Fig.~\ref{fig:se}. The Ni doping has relatively little impact on the low-energy magnetic spectral weight compared with the Cu doping, as represented by the Cu10 sample.

\section{Discussion}

Our measurements of resistivity in Ni and Cu-substituted samples of Fe$_{0.98}$Te$_{0.5}$Se$_{0.5}$ indicate that the TM dopants cause an increase in scattering whose magnitude grows rapidly as the atomic number of the TM dopant deviates from that of Fe. Considering the impact on $T_c$, it appears that even Co has a negative impact.  With increasing atomic number, the $3d$ states of the TM dopants shift to higher binding energies relative to those of Fe.  Density functional calculations of FeSe with various TM dopants have indicated that much of the charge of the extra $3d$ electrons remains close to the TM ions, and it has been proposed that the TM dopants may have their largest impact as scattering centers \cite{PhysRevLett.105.157004}.  Our results are consistent with this proposal.  It is important to note, however, that the TM dopants are not the only source of disorder in our samples.  A diffraction study has found a difference in Fe-Te and Fe-Se bond lengths of 0.15~\AA\ \cite{Tegel2010}, while evidence for short-range segregation of Se and Te has been provided by scanning transmission electron microscopy \cite{Hu2011} and scanning tunneling microscopy \cite{He2011}.  If disorder alone were the key factor, it is not clear why relatively small concentrations of Ni or Cu would have such a large effect on the transport properties.

In the case of BaFe$_2$As$_2$, TM substitution depresses antiferromagnetic order and, above a threshold concentration, induces superconductivity.  It has been proposed that the disorder effect of TM dopants could be sufficient to explain this effect \cite{Vavilov2011}.  In this argument, spin-density-wave order results from scattering of conduction electrons between Fermi surface pockets, and this mechanism is disrupted by impurity scattering to a greater extent than is the electron pairing of the superconducting state.  Our results appear to be inconsistent with this proposal.  We find that the low-energy magnetic spectral weight is not reduced by Ni doping, and actually increases with Cu doping.  Furthermore, the magnetic correlation length is always quite short.  These observations indicate that the important magnetic interactions must be short ranged and are insensitive to the coherence of electronic quasiparticles.

For the sake of completeness, we now want to discuss the possibility of static magnetic order. We know that in the optimally-doped samples, there is no static order, long- or short-ranged, and the spectral weight is concentrated around $(0.5,0.5)$ \cite{qiu:067008,lumsden-2009}. However, with sufficient excess Fe, which suppresses the superconductivity, short-range static order with the wave vector  $(0.5,0,0.5)$, can be induced \cite{xudoping11}. These results indicate that extra Fe stabilizes the bicollinear antiferromagnetic order that is incompatible with superconductivity. Density-functional calculations have confirmed that with excess Fe, the spin configuration changes from collinear with an in-plane ordering wave vector of $(0.5,0.5)$ to a bicollinear structure with in-plane ordering wave vector $(0.5,0)$ \cite{han:067001}.  To see whether similar static magnetic order can be induced with Ni or Cu substitution, we have carried out additional measurements in the $(H0L)$ plane. Those results turn out to be negative. Hence, it seems likely that the Ni dopants substitute for Fe in the lattice.

While the Ni and Cu dopants do not reduce the low-energy magnetic spectral weight, at concentration where no superconductivity is observed, they inhibit magnetic correlations from becoming commensurate at low temperature, and this effect is correlated with the suppression of superconductivity \cite{PhysRevLett.109.227002}. This result has been interpreted as a consequence of the suppressed orbital ordering by the disorder and thus worsened shape matching of the Fermi surface connected by the nesting vector~\cite{PhysRevLett.109.227002}. Our observations that the magnetic excitations in the Ni10 sample peak further away from the commensurate position [Figs.~\ref{fig:meshni7}(c), and \ref{fig:trans}] are consistent with these results.  Thus, the scattering effects of the Ni and Cu dopants might impact orbital as well as charge correlations.

\section{SUMMARY}
In conclusion, our systematic resistivity and inelastic neutron scattering measurements on a series of Fe$_{0.98-z}$Ni$_z$Te$_{0.5}$Se$_{0.5}$ samples have shown that Ni suppresses both the superconductivity and conductivity, with relatively little impact on the magnetic correlations. These effects are weaker than those of the Cu-doped case. We attribute this to the weaker impurity potentials of the Ni dopants. Considering the reports on the substitution-dependent effects in the 122 \cite{PhysRevLett.110.107007,2011arXiv1112.4858B,PhysRevLett.109.167003} and Li(Na)FeAs (111)  \cite{arXiv:1409.5612,PhysRevB.88.245112} systems, it appears that the substitution effects get stronger as the impurity potential of the substituent becomes larger is universal among the Fe-based superconductors. 

\begin{acknowledgments}
We are grateful for stimulating discussions with Weiguo Yin, Xiangang Wan, Alex Frano and Ming Yi. Work at Nanjing University was supported by National Natural Science Foundation of China under contract No.~11374143, Ministry of Education under contract No.~NCET-13-0282, and Fundamental Research Funds for the Central Universities. Work at Lawrence Berkeley National Laboratory and Brookhaven National Laboratory was supported by the Office of Basic Energy Sciences, Division of Materials Science and Engineering, U.S. Department of Energy, under Contract No.~DE-AC02-05CH11231 and DE-AC02-98CH10886, respectively. Research at Oak Ridge National Laboratory's High Flux Isotope Reactor was sponsored by the Division of Scientific User Facilities of the Office of Basic Energy Sciences.
\end{acknowledgments}

%

\end{document}